\documentclass{article}
\usepackage{eurosym}
\usepackage{amssymb}
\usepackage{amsmath}
\usepackage{cite}

\begin{document}

\title{Benney-Lin and Kawahara equations: a detailed study through Lie
symmetries and Painlev\'{e} analysis}
\author{Andronikos Paliathanasis\thanks{%
Email: anpaliat@phys.uoa.gr} \\
{\ \textit{Institute of Systems Science, Durban University of Technology }}\\
{\ \textit{PO Box 1334, Durban 4000, Republic of South Africa}}}
\maketitle

\begin{abstract}
We perform a detailed study on the integrability of the Benney-Lin and
KdV-Kawahara equations by using the Lie symmetry analysis and the
singularity analysis. We find that the equations under our consideration
admit integrable travelling-wave solutions. The singularity analysis is
applied for the partial differential equations and the generic algebraic
solution is presented.

\end{abstract}

\section{Introduction}

We investigate the integrability of the Benney-Lin partial differential
equation (PDE) \cite{ben1,ben2}
\begin{equation}
u_{t}+uu_{x}+u_{xxx}+\beta \left( u_{xx}+u_{xxxx}\right) +\alpha u_{xxxxx}=0
\label{bl.01}
\end{equation}%
with $\alpha \neq 0,$ by applying the theory of symmetries of differential
equations and Painlev\'{e} analysis.

The Benney-Lin equation follows from the theory of fluid dynamics and more
specifically describes the propagation of long waves on thin film. For $%
\beta =0$, the equation becomes the Kawahara equation \cite{kawa} while when
$\alpha =0$ equation (\ref{bl.01}) takes the form of the derivative of the
KdV equation. The Benney-Lin equation has been widely studied in the
literature \cite{ben3,ben4,ben5,ben6,ben7,ben8,ben9}. However, in this work
we prove the integrability of the Benney-Lin equation and of the Kawahara
equation for the similarity solutions by applying the theory of Lie
symmetries and the singularity analysis.

The theory of symmetries for differential equations it is a powerful
mathematical tool for the study of nonlinear differential equations, and
consequently of the study of nonlinear phenomena, for instance see \cite%
{ls01,ls02,ls03,ls04,ls05,ls06,ls07} and references therein. The novelty of
Lie symmetries is that they provide invariant surfaces which can be used to
simplify the nonlinear differential equation or to determine conservation
laws which are necessary for the integration of differential equations.

Singularity analysis is another major tool for the study of the
integrability of differential equations. In contrary to the symmetry
analysis, singularity analysis is based on the existence of movable
singularities in the differential equation. Singularity analysis is
associated with the French school led by Painlev\'{e} in the last years of
the nineteenth century and the early years of the twentieth century \cite%
{pan2,pan3,pan4} which was actually inspired by the successful application
to the determination of the third integrable case of Euler's equations for a
spinning top by Kowalevskaya \cite{kowa}. Modern applications of the
singularity analysis in nonlinear analysis are presented in \cite%
{sin1,sin2,sin3,sin4} and references therein.

In this work investigate the integrability of the Benney-Lin and Kawahara
equations for the PDEs by using the Lie point symmetries and the singularity
analysis. Specifically we prove that similarity transformations provide
integrable ordinary differential equations, while the two PDEs of our
consideration passes the singularity analysis. The plan of the paper is as
follows.

The main properties and definitions of the mathematical tools that we apply
in this work are presented in Section \ref{sec2}. In Section \ref{sec3}, we
present the detailed Lie symmetry analysis for the Benney-Lin equation. We
find that the Benney-Lin equation admits three Lie point symmetries which
form the $A_{3,1}$ Lie algebra. The seven possible reductions of the
Benney-Lin equation with the use of the Lie invariants are studied. The
resulting differential equations are studied again in terms of symmetries,
or with the use of singularity analysis. The latter study is performed in
Section \ref{sec4}. More specifically we find that the ordinary differential
equations (ODEs) which follow from the similarity transformations of the
Benney-Lin equations pass the singularity test, only after we apply firstly
a recently proposed technique in the theory of singularity analysis. In
Section \ref{sec5}, we extend our analysis in the case of a generalized
Benney-Lin equation. In Section \ref{sec6} we apply the Painlev\'{e}
analysis directly for the PDEs without considering any reduction with the
Lie symmetries. We find that the Benney-Lin equation and the generalized
Benney-Lin equations pass the Painlev\'{e} test. The generic solution is
given in a Right Painlev\'{e} Series. Our discussion and conclusions are
presented in Section \ref{sec7}.

\section{Preliminaries}

\label{sec2}

In this section, we present the basic mathematical material and definitions
which we use in this work. More specifically we discuss the theory of
symmetries for differential equations and the singularity analysis.

\subsection{Lie symmetries of differential equations}

We briefly discuss the theory of Lie symmetries of differential equations
and the application of the differential invariants for the construction of
similarity solutions.

Let $\Phi $ describe the map of an one-parameter point transformation such
as $u^{\prime }\left( t^{\prime },x^{\prime }\right) =\Phi \left( u\left(
t,x\right) ;\varepsilon \right) \ $with infinitesimal transformation%
\begin{eqnarray}
t^{\prime } &=&t+\varepsilon \xi ^{t}\left( t,x,u\right)  \label{sv.12} \\
x^{\prime } &=&x^{i}+\varepsilon \xi ^{x}\left( t,x,u\right) \\
u &=&u+\varepsilon \eta \left( t,x,u\right)  \label{sv.13}
\end{eqnarray}%
and generator
\begin{equation}
X=\frac{\partial t^{\prime }}{\partial \varepsilon }\partial _{t}+\frac{%
\partial x^{\prime }}{\partial \varepsilon }\partial _{x}+\frac{\partial
u^{\prime }}{\partial \varepsilon }\partial _{u},  \label{sv.16}
\end{equation}%
where~{$\varepsilon $ is an infinitesimal parameter}. \

Consider function $u\left( x^{i}\right) $ to be a solution of the
differential equation
\begin{equation}
\mathcal{H}\left( u,u_{t},u_{x},u_{tt},u_{xx},u_{tx},u_{xxx},...\right) =0;
\end{equation}%
then under the map $\Phi $, function $u^{\prime }\left( x^{i\prime }\right)
=\Phi \left( u\left( x^{i}\right) \right) $ remains a solution for the
differential equation iff the differential equation $\mathcal{H\,}=0$
remains invariant under the action of the map $\Phi $, that is
\begin{equation}
\Phi \left( \mathcal{H}\left(
u,u_{t},u_{x},u_{tt},u_{xx},u_{tx},u_{xxx},...\right) \right) =0.
\end{equation}%
The vector field $X$ is called a Lie point symmetry for the differential
equation $\mathcal{H\,}=0$.

The mathematical formulation of the latter is expressed by the condition
\begin{equation}
X^{\left[ n\right] }\left( \mathcal{H}\right) =0,  \label{sv.17}
\end{equation}%
where $X^{\left[ n\right] }$ describes the $n$th prolongation/extension of
the symmetry vector in the jet-space of variables $\left\{
u,u_{t},u_{x},...\right\} $ defined as%
\begin{equation*}
X^{\left[ n\right] }=X+\eta _{i}^{\left[ 1\right] }\partial
_{u_{i}}+...+\eta ^{\left[ n\right] }\partial _{u_{i_{i}i_{j}...i_{n}}},
\end{equation*}%
where $u_{i}=\frac{\partial u}{\partial y^{i}},~y^{i}=\left( t,x\right) $ and%
\begin{equation}
\eta _{i}^{\left[ n\right] }=D_{i}\eta ^{\left[ n-1\right]
}-u_{ii_{2}...i_{n-1}}D_{j}\left( \xi ^{j}\right) ~,~i\succeq 1.
\end{equation}

One of the novelties of the existence of a Lie point symmetry is that
invariant functions can be determined. If $X$ is an admitted Lie point
symmetry, the solution of the associated Lagrange's system,%
\begin{equation}
\frac{dt}{\xi ^{t}}=\frac{dx}{\xi ^{x}}=\frac{du}{\eta },
\end{equation}%
provides the zeroth-order invariants,~$U^{A\left[ 0\right] }\left(
t,x,u\right) $ which are applied to reduce the number of independent
variables in PDEs or reduce the order for ODE.

When the admitted symmetries of a given differential equation is sufficient
to reduce the differential equation into a first-order ODE, or into
well-known integrable differential equation we say that the given
differential equation is integrable in terms of Lie symmetries. However,
symmetry is not the unique method to study the integrability of differential
equations. Another major method for the study of the integrability of
differential equations is presented in the following lines.

For more details on the symmetry analysis of differential equations we refer
the reader to the standard references \cite{book1,book2,book3}, while a
review on the application of Lie symmetries to the theory of turbulence can
be found in \cite{bk04}.

\subsection{Singularity analysis}

An alternate tool to study the integrability of differential equations is
the Painlev\'{e} analysis, also known as singularity analysis. The modern
treatment of singularity analysis is summarized in the ARS algorithm \cite%
{Abl1,Abl2,Abl3}. ARS stands for Ablowitz, Ramani and Segur.

There are three main steps to the algorithm which summarized as: (a)
determine the leading-order term, and show if a moveable singularity exists,
(b) determine the resonances which indicates the position of the constants
of integration and (c) write a Painlev\'{e} Series with exponent and step as
determined in the previous step and study if it is a solution of the
original differential equation, the latter it is called the consistency test.

There are various criteria which should be satisfied for the values of the
exponent of the leading-order term or for the values of the resonances these
are summarized in the review article of Ramani et al. \cite{buntis}. In the
following line we discuss in details the three steps of the ARS algorithm.

Let $H\left( x,y\left( x\right) ,y^{\prime }\left( x\right) ,...\right) =0$
describe an ODE, where obviously $x$ denotes the independent variable and $%
y\left( x\right) $ the dependent variable. In order to determine the
leading-order term we substitute $y=a(x-x_{0})^{p}$ into the differential
equation, where $x_{0}$ is the location of the \ movable singularity. The
dominant powers are selected in order that the terms to share a common
scaling symmetry. In order for the singularity to be a pole, the exponent $p$
\ should take negative integer values. However, $p$ can also take fractional
exponents, even positive ones as the repeated differentiation of a positive
fractional exponent eventually gives a negative exponent and so a
singularity.

The resonances are determined by substituting $%
y=a(x-x_{0})^{p}+m(x-x_{0})^{p+r} $ into the differential equation for given
$a $ and $p$. The dominant terms linear in $m$ provide a polynomial in $r,$
the zeros of which are the resonances. One of the zeros should be $-1$,
which is important for the existence of the singularity. The acceptable
values of the resonances are rational numbers. When the rest of the
resonances are positive, the possible solution may be written by a Right
Painlev\'{e} Series (Laurent expansion), when the resonances are negative,
the possible solution is expressed by a Left Painlev\'{e} Series, while,
when there are positive and negative resonances, the possible solution is
expressed by a Mixed Painlev\'{e} Series (full Laurent expansion).

Finally, in the consistency test, the constants of integration are
determined while the coefficients of the Laurent expansions are derived,
which provide the exact form of the algebraic solution for the given
differential equation.

\section{Point symmetries of the Benney-Lin and Kawahara equations}

\label{sec3}

We apply Lie's theory for the Benney-Lin equation (\ref{bl.01}) for $\beta
\neq 0$ and $\beta =0$ and we derive the following Lie point symmetries%
\begin{equation*}
X_{1}=\partial _{t}~,~X_{2}=\partial _{x}~~X_{3}=t\partial _{x}+\partial _{u}
\end{equation*}%
with Lie Brackets presented in Table \ref{table1}. We infer that these three
vector fields form the $A_{3,1}$ algebra in the Morozov-Mubarakzyanov
classification scheme \cite{mor1,mor2,mor3,mor4}. The admitted Lie point
symmetries are fewer than the number of symmetries required to reduce the
fifth-order PDE into a first-order ODE. However, for every possible
reduction we search again for the admitted Lie point symmetries. In the
following we remark when there are different results for the special case
with $\beta =0$.

\begin{table}[tbp] \centering%
\caption{Lie Brackets of the Lie point symmetries of the Benney-Lin equation}%
\begin{tabular}{c|ccc}
\hline\hline
$\left[ ~\mathbf{,}~\right] $ & $\mathbf{X}_{1}$ & $\mathbf{X}_{2}$ & $%
\mathbf{X}_{3}$ \\ \hline
$\mathbf{X}_{1}$ & $0$ & $0$ & $X_{2}$ \\
$\mathbf{X}_{2}$ & $0$ & $0$ & $0$ \\
$\mathbf{X}_{3}$ & $-X_{2}$ & $0$ & $0$ \\ \hline\hline
\end{tabular}%
\label{table1}%
\end{table}%

We continue by calculating the adjoint representation for the Lie algebra in
order to determine the one-dimensional optimal system. For every element of
the\ admitted Lie algebra, the adjoint representation is defined as \cite%
{olverb}
\begin{equation}
Ad\left( \exp \left( \varepsilon X_{i}\right) \right)
X_{j}=X_{j}-\varepsilon \left[ X_{i},X_{j}\right] +\frac{1}{2}\varepsilon
^{2}\left[ X_{i},\left[ X_{i},X_{j}\right] \right] +...~\text{.}
\label{ad.01}
\end{equation}%
In Table \ref{tabl2} the adjoint representation for the Lie symmetries of
the Benney-Lin equation is presented. \ Hence from Table \ref{tabl2} the
one-dimensional optimal systems are determined to be%
\begin{equation*}
X_{1},~X_{2}~,~X_{3},~X_{1}+cX_{2}~\text{and }X_{1}+\gamma X_{3}\text{.}
\end{equation*}

\begin{table}[tbp] \centering%
\caption{Adjoint representation for the Lie point symmetries of the
Benney-Lin equation}%
\begin{tabular}{c|ccc}
\hline\hline
$Ad\left( \exp \left( \varepsilon X_{i}\right) \right) X_{j}$ & $X_{1}$ & $%
X_{2}$ & $X_{3}$ \\ \hline
$X_{1}$ & $X_{1}$ & $X_{2}$ & $-\varepsilon X_{2}+X_{3}$ \\
$X_{2}$ & $X_{1}$ & $X_{2}$ & $X_{3}$ \\
$X_{3}$ & $X_{1}+\varepsilon X_{2}$ & $X_{2}$ & $X_{3}$ \\ \hline\hline
\end{tabular}%
\label{tabl2}%
\end{table}%

\subsection{Reduction by use of Lie invariants}

From the admitted Lie point symmetries, the possible reductions that we can
perform are \ with the use of the symmetry vectors: (i) $X_{1},~$(ii) $X_{2}$%
, (iii) $X_{3}$, (iv)~$X_{1}+cX_{2},~$(v)$~X_{1}+\gamma X_{3}$.~

Reduction (i) provides the static solution, reduction (ii) provides the
stationary solution, reduction (iii) is a scaling solution, while reduction
(iv) provides the travelling-wave solution which is of special interest. The
constant $c$ denotes the wave speed. The remaining reductions provide
combinations of the previously discussed solutions.

Moreover, reductions (ii), (iii) and (vi) provide a linear first-order ODE
as reduced equation, while a nonlinear fifth-order ODE is provided in the
reductions (i), (iv), (v) and (vii).

\subsubsection{Reductions (ii)~and~(iii)}

The Lie invariants which correspond to the Lie point symmetry $\gamma
_{1}X_{2}+\gamma _{2}X_{3}$~are
\begin{equation}
t~\ \text{and }u=\frac{\gamma _{2}x}{\gamma _{2}t+\gamma _{1}}+v.
\label{bl.02}
\end{equation}%
When we consider $t$ to be the new independent variable and $v$ the new
dependent variable, equation (\ref{bl.01}) becomes%
\begin{equation}
\left( \gamma _{1}+\gamma _{2}t\right) v_{t}+\gamma _{2}v=0,  \label{bl.03}
\end{equation}%
with exact solutions%
\begin{eqnarray}
v\left( t\right)  &=&\frac{v_{0}}{\gamma _{2}t+\gamma _{1}}~,~\gamma
_{2}\neq 0,  \label{bl.04} \\
v\left( t\right)  &=&v_{0}~,~\gamma _{2}=0.  \label{bl.05}
\end{eqnarray}

\subsubsection{Reduction (i) and (iv)}

From the symmetry vector $X_{1}+cX_{2}$ we calculate the Lie invariants%
\begin{equation}
s=x-ct~,~u~.  \label{bl.06}
\end{equation}

Now, when we consider $s$ to be the new independent variable and $u=u\left(
s\right) $ the dependent variable, the Benney-Lin equation, (\ref{bl.01}),
is reduced to the fifth-order ODE
\begin{equation}
\alpha u^{\prime \prime \prime \prime \prime }+\beta \left( u^{\prime \prime
\prime \prime }+u^{\prime \prime }\right) +u^{\prime \prime \prime
}+u^{\prime }\left( u-c\right) =0,  \label{bl.07}
\end{equation}%
which provides the travelling-wave solution or the stationary solution for $%
c=0$, and prime denotes total derivative with respect to $s$, i.e. $%
u^{\prime }=\frac{du}{ds}$.

Easily we observe that any travelling-wave solution $u\left( s\right) $ of
equation (\ref{bl.07}) describes a stationary solution $v\left( s\right) $,
where $v\left( s\right) =u\left( s\right) -c$. Moreover, equation (\ref%
{bl.07}) in the new variable $v\left( s\right) $ can be written as follows%
\begin{equation}
\left( av^{\prime \prime \prime \prime }+\beta \left( v^{\prime \prime
\prime }+v^{\prime }\right) +v^{\prime \prime }+\frac{v^{2}}{2}\right)
^{^{\prime }}=0,  \label{bl.08a}
\end{equation}%
or equivalently
\begin{equation}
av^{\prime \prime \prime \prime }+\beta \left( v^{\prime \prime \prime
}+v^{\prime }\right) +v^{\prime \prime }+\frac{v^{2}}{2}=\delta ,
\label{bl.08}
\end{equation}%
which is a fourth-order ODE.

Equation (\ref{bl.08}) admits only the autonomous Lie point symmetry $\bar{X}%
=\partial _{s}\,,~$\ which is an inherited symmetry. The latter symmetry
vector is applied and equation (\ref{bl.08}) is written in the form of the
following third-order nonlinear differential equation%
\begin{equation}
a\frac{d}{dz}\left( y\left( \frac{dy}{dz}\right) ^{2}+2y^{3}\left( \frac{%
d^{2}y}{dz^{2}}\right) \right) +2\beta y\left( \left( \frac{d^{2}y}{dz^{2}}%
\right) +y^{2}\left( \frac{dy}{dz}\right) ^{2}+1\right) +2\frac{d}{dz}\left(
y^{2}\right) +z^{2}-2\delta =0,  \label{bl.09}
\end{equation}%
where $z=v\left( s\right) $ and \thinspace $y=v^{\prime }$. This third-order
ODE does not admit Lie point symmetries. Hence other method should be
considered to study the integrability of this ODE. In our study we consider
the singularity analysis. However, before we proceed with such an analysis
we observe that for the Kawahara equation expression (\ref{bl.09}) is a
total derivative, that is, $y\left( z\right) $ is a solution of the
second-order ODE%
\begin{equation}
y\left( \frac{dy}{dz}\right) ^{2}+2y^{3}\left( \frac{d^{2}y}{dz^{2}}\right)
+2y^{2}+\frac{z^{3}}{3}-2\delta z+\kappa =0.  \label{bl.10}
\end{equation}%
This equation admits no Lie point symmetries. However, it can be integrated
as follows%
\begin{equation}
e^{-\frac{1}{2y^{2}}}\frac{dy}{dz}+\frac{2}{3}y^{3}+\frac{z^{4}}{12}-\delta
z^{2}+\kappa z+\lambda =0.  \label{bl.11}
\end{equation}

Before we study equation, (\ref{bl.08}), with the use of singularity
analysis we study the reductions (v) and (vii).

\subsubsection{Reduction (v)}

The zeroth-order invariants of the Lie symmetry vector $X_{1}+cX_{2}+\gamma
X_{3}$ are%
\begin{equation}
s=\left( x-ct\right) -\frac{\gamma }{2}t^{2}~,~u=\gamma t+w.  \label{bl.12}
\end{equation}

Consequently, in the new variables, the Benney-Lin equation is simplified as%
\begin{equation}
\alpha w^{\prime \prime \prime \prime \prime }+\beta \left( w^{\prime \prime
\prime \prime }+w^{\prime \prime }\right) +w^{\prime \prime \prime
}+w^{\prime }\left( w-c\right) +\gamma =0  \label{bl.13}
\end{equation}%
which reduces to (\ref{bl.07}) for $\gamma =0$. In a similar way, the
reduced equation (\ref{bl.13}) can be written as
\begin{equation}
\alpha v^{\prime \prime \prime \prime \prime }+\beta \left( v^{\prime \prime
\prime \prime }+v^{\prime \prime }\right) +v^{\prime \prime \prime
}+v^{\prime }v+\gamma =0  \label{bl.14}
\end{equation}%
in which $v\left( s\right) =w\left( s\right) -c$. \ By following the steps
for the travelling-wave reduction, this fifth-order equation can be reduced
to a third-order equation, or a first-order equation when $\beta =0$.

Equation (\ref{bl.14}) provides the fourth-order equation%
\begin{equation}
av^{\prime \prime \prime \prime }+\beta \left( v^{\prime \prime \prime
}+v^{\prime }\right) +v^{\prime \prime }+\frac{v^{2}}{2}+\gamma s=\delta
\label{bl.15}
\end{equation}%
which does not admit any symmetry for $\gamma \neq 0$ and cannot be reduced
to a third-order as can be equation (\ref{bl.08}).

In the following Section, we apply the singularity analysis to study the
integrability for the travelling-wave solutions (\ref{bl.08}) and (\ref%
{bl.15}). \

\section{Singularity analysis}

\label{sec4}

We proceed our analysis by applying the ARS algorithm for the fourth-order
ODE (\ref{bl.08}) which provides the travelling-wave solution.

We substitute $v\left( s\right) =v_{0}s^{p}$ into (\ref{bl.08}) and obtain
\begin{equation}
ap\left( p-1\right) \left( p-2\right) \left( p-3\right) s^{p-4}+p\beta
\left( \left( p-1\right) \left( p-2\right) s^{p-3}+s^{p-1}\right) +p\left(
p-1\right) s^{p-2}+\frac{v_{0}}{2}s^{2p}-\frac{\delta }{v_{0}}=0.
\label{bl.16}
\end{equation}%
Consequently from this expression it follows necessarily that~the
leading-order terms are $p-4=2p$, which means that $p=-4$. Hence, in order
that the leading-order terms cancel coefficient $v_{0}$ has the value $%
v_{0}=-1680\alpha $.

The second step is to determine the resonances. We replace
\begin{equation}
v\left( s\right) =-1680\alpha s^{-4}+ms^{-4+r}
\end{equation}%
in (\ref{bl.08}) and we linearize around the $m\rightarrow 0$, the
requirement that the coefficient of the leading-order terms be zero provides
the algebraic equation%
\begin{equation}
\left( r+1\right) \left( r-12\right) \left( r^{2}-11r+70\right) =0,
\label{bl.17}
\end{equation}%
with solutions%
\begin{equation}
r_{1}=-1,~r_{2}=12,~r_{3}=\frac{11-i\sqrt{159}}{2}~,~r_{4}=\frac{11+i\sqrt{%
159}}{2}.  \label{bl.18}
\end{equation}

Because two of the resonances are imaginary we conclude that equation (\ref%
{bl.08}) does not pass the Painlev\'{e} test. Hence no conclusion for the
integrability of the equation can be made.

Recently in \cite{sinandronikos} a new technique proposed on the ARS
algorithm in order to extend the application range of the ARS algorithm. The
idea is based that the movable singularity in finite time where the ARS
algorithm required can be coordinate dependent.

Indeed from (\ref{bl.16}) it is clear that the exponent $p-4$ can be a
leading term for $p=1,$~$p=2$~or $p=3$ . However, because $p>0,~$we do not
have any movable singularity. That it is not true for the equation which
follows for the the coordinate transformation $V\left( s\right) =\left(
v\left( s\right) \right) ^{-1}$ in (\ref{bl.08}).

By replacing $V\left( s\right) =V_{0}s^{p}$ in the new equation we find the
following exponents and corresponding coefficients
\begin{equation*}
p_{1}=-1~,~p_{2}=-2~,~p_{3}=-3~\text{for arbitrary }V_{0}\text{.}
\end{equation*}%
\begin{equation*}
p_{4}=-4~~,~V_{0}^{1}=24\frac{\alpha }{\delta }~,~\delta \neq 0,
\end{equation*}

We apply the second step of the ARS algorithm and the resonances are
determined to be%
\begin{eqnarray*}
p_{1} &\rightarrow &-1:r_{1}=-1~,~r_{2}=0~,~r_{3}=1~,~r_{4}=2, \\
p_{2} &\rightarrow &-2:r_{1}=-1~,~r_{2}=0~,~r_{3}=1~,~r_{4}=-2, \\
p_{3} &\rightarrow &-3:r_{1}=-1~,~r_{2}=0~,~r_{3}=-1~,~r_{4}=-2, \\
p_{4} &\rightarrow &-4:r_{1}=-1~,~r_{2}=-2~,~r_{3}=-3~,~r_{4}=-4.
\end{eqnarray*}

There are four possible solutions. Solution with $p_{1}$ is expressed by a
Right Painlev\'{e} Series, solutions with exponent $p_{3}$ and $p_{4}$ by
Left Painlev\'{e} Series, while the solution with exponent $p_{2}$ is
expressed by a Mixed Painlev\'{e} Series. Before we conclude about the
validity of the solutions, the consistency test should be performed.

For the exponent $p_{1}$ the solution is written as
\begin{equation}
V\left( s\right) =V_{0}s^{-1}+V_{1}+V_{2}s+\sum\limits_{I=3}^{\infty
}V_{I}s^{-1+I},
\end{equation}%
where from the consistency test we find that $V_{0},~V_{1}$ and $V_{2}$ are
arbitrary while the rest coefficients are expressed as functions
\begin{equation}
V_{I}=V_{I}\left( V_{0},V_{1},V_{2}\right)
\end{equation}%
with first coefficient~$V_{3}$ to be%
\begin{equation}
V_{3}=-\frac{\delta V_{0}^{4}+24\alpha V_{1}^{3}+2V_{0}^{2}V_{1}+\beta
\left( 6V_{0}^{3}V_{2}-V_{0}^{2}-V_{0}V_{1}^{2}\right) -48V_{0}V_{1}V_{2}}{%
24\alpha V_{0}^{2}}.
\end{equation}

For the two Left Painlev\'{e} Series, and the Mixed Painlev\'{e} Series, the
consistency test fails. Nevertheless, we conclude that the fourth-order ODE (%
\ref{bl.08}) is integrable according to the singularity analysis. We proceed
by studying if the fifth-order ODE (\ref{bl.14}) passes the Painlev\'{e}
test.

For equation (\ref{bl.14}) \ in order the ARS algorithm to succeed we apply
first the coordinate transformation $v\left( s\right) \rightarrow
V^{-1}\left( s\right) $, from where we find the possible exponents to be%
\begin{equation*}
p_{1}=-1~,~p_{2}=-2~,~p_{3}=-3~,~p_{4}=-4\text{ and }p_{5}=-5.
\end{equation*}%
From these exponents only the solution correspond to the $p_{1}$ passes all
the steps of the ARS algorithm with resonances
\begin{equation*}
r_{1}=-1,~r_{2}=0~,~r_{3}=1~,~r_{4}=2\text{ and }r_{5}=3
\end{equation*}%
and solution given by a Right Painlev\'{e} Series with step one and the
first four coefficients as integration constants. We remark that the
analysis is valid for arbitrary parameter $\beta $.

We conclude that there exists similarity solutions for the Benney-Lin
equation (\ref{bl.01}). We continue by extending our analysis for a
generalization of the Benney-Lin equation.

\section{The generalized Benney-Lin equation}

\label{sec5}

The generalized Benney-Lin and generalized Kawahara equation proposed in
\cite{gbin} is
\begin{equation}
u_{t}+u^{n}u_{x}+u_{xxx}+\beta \left( u_{xx}+u_{xxxx}\right) +\alpha
u_{xxxxx}=0~,  \label{gbn.01}
\end{equation}%
with$~\alpha \neq 0~,~n\neq 1,2$.

We apply the symmetry analysis and we find that equation (\ref{gbn.01}) is
invariant under the two-dimensional Lie algebra $2A_{1}~$comprised of the
symmetry vector fields $X_{1}$ and $X_{2}$. The application of the Lie
invariants (\ref{bl.06}) derived by the generic Lie symmetry $X_{1}+cX_{2}$
provide the fifth-order ODE%
\begin{equation}
\alpha u^{\prime \prime \prime \prime \prime }+\beta \left( u^{\prime \prime
\prime \prime }+u^{\prime \prime }\right) +u^{\prime \prime \prime
}+u^{\prime }\left( u^{n}-c\right) =0,  \label{gbn.02}
\end{equation}%
which can be integrated as%
\begin{equation}
\alpha u^{\prime \prime \prime \prime }+\beta \left( u^{\prime \prime \prime
}+u^{\prime }\right) +u^{\prime \prime }+\left( \frac{1}{n+1}%
u^{n+1}-cu\right) =\delta .  \label{gbn.03}
\end{equation}

In a similar way as with (\ref{bl.01}) the generalized Benney-Lin equation
is reduced to the third-order ode%
\begin{equation}
a\frac{d}{dz}\left( \frac{y}{2}\left( \frac{dy}{dz}\right) ^{2}+y^{3}\left(
\frac{d^{2}y}{dz^{2}}\right) \right) +\beta y\left( \left( \frac{d^{2}y}{%
dz^{2}}\right) +y^{2}\left( \frac{dy}{dz}\right) ^{2}+1\right) +\frac{d}{dz}%
\left( y^{2}\right) +\left( \frac{z^{n+1}}{n+1}-cz\right) -\delta =0,
\label{gbn.04}
\end{equation}%
or to the first-order ODE for $\beta =0$,%
\begin{equation}
e^{-\frac{1}{2y^{2}}}\frac{dy}{dz}+\frac{2}{3}y^{3}+\frac{z^{n+3}}{2\left(
n+2\right) \left( n+1\right) }-c\frac{z^{3}}{6}-\delta z^{2}+\kappa
z+\lambda =0,  \label{gbn.05}
\end{equation}%
in which $z=v\left( s\right) $ and \thinspace $y=v^{\prime }$ and $\delta
,~\kappa ,~\lambda $ are constants of integration. \ We are interested in
the application of the singularity analysis for equation (\ref{gbn.03}) in
order to study the integrability of the generalized Benney-Lin equation for $%
\beta \neq 0$. Hence, we continue with the application of the ARS algorithm.

We substitute $u\left( s\right) =u_{0}s^{p}$ and we find that the only
dominant term has exponent $p=-\frac{4}{n}$ for $n>0,$ while for the
coefficient $u_{0}$ we find
\begin{equation}
u_{0}=\left( -1\right) ^{\frac{1}{n}}\left( \frac{\alpha \left( 4+3n\right)
\left( 2+n\right) \left( n+1\right) }{8n^{4}}\right) ^{\frac{1}{n}}.
\label{gbn.06}
\end{equation}

The second step of the ARS provides the resonances
\begin{equation}
r_{1}=-1~,~r_{2}=\frac{4}{n}\left( n+2\right) ~,~r_{3,4}=\frac{3n+8\pm \sqrt{%
-\left( 15n^{2}+80n+64\right) }}{2n}.  \label{gbn.07}
\end{equation}%
Because for $n>0,$ it holds $\left( 15n^{2}+80n+64\right) >0$, resonances $%
r_{3}$ and $r_{4}$ are complex which means that (\ref{gbn.03}) fails to pass
the Painlev\'{e} test. Note that for $n=1$ resonances (\ref{bl.18}) are
recovered.

We follow the same procedure with the previous section, i.e. we perform the
change of variable $u\rightarrow U^{-1}$ and we find that the Right Painlev%
\'{e} Series%
\begin{equation}
U\left( s\right) =U_{0}s^{-1}+U_{1}+U_{2}s+\sum\limits_{I=3}^{\infty
}U_{I}s^{-1+I}  \label{gbn.08}
\end{equation}%
is a solution of equation (\ref{gbn.03}), where $U_{0},~U_{1}$ and $U_{2}$
are constants of integration and the remaining coefficient constants are
expressed as $U_{I}=U_{I}\left( U_{0},U_{1},U_{2},n,\alpha ,\beta ,c\right) $%
.

It is important to mention the first step of the ARS algorithm provides more
exponents $p$ for the differential equation (\ref{gbn.03}) under the
transformation $u\rightarrow U^{-1}$. In a similar way with the case $n=1$.
However, the Left Painlev\'{e} Series and the Mixed Painlev\'{e} Series fail
the consistency test.

Thereafter, the generalized Benney-Lin equation passes the Painlev\'{e} test
and it is integrable for any allowed value of the parameter $n$.

\section{Singularity analysis for the Benney-Lin equation}

\label{sec6}

So far the Painlev\'{e} analysis we have performed for the Benney-Lin
equation was based on the reduction of the equation by use of the Lie
invariant and perform the singularity analysis for the resulting ODE.
Nevertheless it is possible to perform the singularity analysis directly for
the\ PDE (\ref{bl.01}). The application of Painlev\'{e} analysis for PDEs is
summarized in \cite{ptpde1,ptpde2,ptpde3}.

In the case of PDEs the main steps of the ARS remain the same. We
demonstrate the main steps in the following well-studied example.

We consider the nonlinear evolution equation \cite{ptpde3}%
\begin{equation}
uu_{xx}+\left( u_{x}\right) ^{2}-u^{2}u_{t}=0  \label{gbn.09}
\end{equation}
and we substitute $u=u_{0}\phi \left( t,x\right) ^{p}.$ We follow the first
step of the ARS algorithm as explained in Section \ref{sec2}. We find the
exponent $p=-1$, with $u_{0}=-\frac{u_{x}^{2}}{u_{t}}.$\

The resonances are determined by considering the substitution%
\begin{equation}
u=u_{0}\phi \left( t,x\right) ^{-1}+m\phi \left( t,x\right) ^{-1+r},
\label{gbn.10}
\end{equation}%
in (\ref{gbn.09}) where again the requirement the leading-order terms which
are linear in $m$ to be zero gives the algebraic equation $\left( r+1\right)
\left( r-1\right) =0$, i.e. $r=-1$ and $r=1$.

The main difference of the algorithm between ODEs and PDEs is in the
consistency test. To perform the consistency test we replace in (\ref{gbn.09}%
) the following expression%
\begin{equation}
u\left( t,x\right) =u_{0}\left( t,x\right) \phi \left( t,x\right)
^{-1}+u_{1}\left( t,x\right) +\sum\limits_{I=2}^{\infty }u_{I}\left(
t,x\right) \phi \left( t,x\right) ^{-1+I},
\end{equation}%
where the first two coefficients in powers of $\phi \left( t,x\right) $
provide the following Painlev\'{e}-Backlund equations (PB) \cite{ptpde1}
\begin{equation}
\left( \phi _{x}\right) ^{2}+u_{0}\phi _{t}=0,  \label{pp.01}
\end{equation}%
\begin{equation}
u_{0}\left( \phi _{xx}+u_{0x}\right) -2u_{1}\left( \phi _{x}^{2}+u_{0}\phi
_{t}\right) =0.  \label{pp.02}
\end{equation}%
The solution of these equations provides the functional form of the
coefficient functions. In general an Ansatz should be made for the function $%
\phi \left( t,x\right) $, one of the most well-known Ansatze is that of
Conte \cite{ptpde1} where%
\begin{equation}
\phi \left( t,x\right) =\frac{\alpha +\beta e^{k\left( x-ct\right) }}{\gamma
+\delta e^{k\left( x-ct\right) }}.  \label{pp.03}
\end{equation}%
For this Ansatz we find that $u_{0}\left( t,x\right) =-kc^{2}\frac{\alpha
+\beta e^{k\left( x-ct\right) }}{\left( \gamma +\delta e^{k\left(
x-ct\right) }\right) ^{2}},$ while $u_{1}\left( t,x\right) $ is arbitrary.

For the Benney-Lin equation we perform the change of variable $%
u=U^{-1}\left( t,x\right) $ and for the new variable we select $U\left(
t,x\right) =U_{0}\phi ^{p}\left( t,x\right) $. For the exponent $p$ we find
that only the Right Laurent expansion for $p=-1$ \ satisfies the consistency
test. In particular the solution is given by
\begin{equation}
U\left( t,x\right) =U_{0}\left( t,x\right) \phi ^{-1}\left( t,x\right)
+U_{1}\left( t,x\right) +\sum\limits_{I=2}^{\infty }U_{I}\left( t,x\right)
\phi \left( t,x\right) ^{-1+I},
\end{equation}%
where $U_{0},~U_{1},~U_{2}$ and $U_{4}$ are found to be arbitrary functions
and $U_{I}\left( t,x\right) =U_{I}\left( U_{0}\left( t,x\right) ,U_{1}\left(
t,x\right) ,U_{2}\left( t,x\right) ,U_{4}\left( t,x\right) \right) $. The
same result is found for the generalized Benney-Lin equation (\ref{gbn.01}).

\section{Conclusions}

\label{sec7}In this work we applied two major mathematical tools for the
study of the integrability of the Benney-Lin and the KdV-Kawahara equations.
In particular, we applied the theory of symmetries of differential equations
and the singularity analysis.

The symmetry analysis provided that the Benney-Lin equation is invariant
under a three-dimensional Lie algebra, the $A_{3,1}$. We considered the
seven possible reductions by using the corresponding zeroth-order
invariants. The resulting ODEs have been solved either by applying Lie
theory or by considering the ARS algorithm for ODEs. In order for the
similarity solutions to be determined with the use of the singularity
analysis a change a variable had to be performed. We concluded that the
Benney-Lin equation is integrable according to the similarity solutions
given by the Lie symmetries. The analysis extended and for the case of the
generalized Benney-Lin equation where we ended up with the same conclusion
on the integrability.

Finally, we studied the PDE by using the singularity analysis where we find
that the Benney-Lin equations passes the Painlev\'{e} test and is integrable.


\begin{thebibliography}{99}
\bibitem{ben1} D.J. Benney, J.\ Math. Phys. 45, 150 (1966)

\bibitem{ben2} S.P. Lin, J. Fluid. Mech. 62, 417 (1974)

\bibitem{kawa} T. Kawahara, J. Phys. Soc. Japan, 33, 260 (1972)

\bibitem{ben3} H.A. Bagioni and F. Linares, J. Math. Anal. Appl. 211, 131
(1997)

\bibitem{ben4} M. Sepulveda and O. Vera, Proc. Appl. Math. Mech. 7, 2020033
(2007)

\bibitem{ben5} H. Holden, C. Lubich and N.H. Risebro, Math. Comp. 82, 173
(2013)

\bibitem{ben6} D. Kaya and K. Al-Khaled, Phys. Lett. A 365, 433 (2007)

\bibitem{ben7} T. Pinguello de Andrade, F.\ Cristofani and F. Natali, J.
Math. Phys. 58, 051504 (2017)

\bibitem{ben8} A.H. Badali, M.S. Hashemi and M.\ Ghahremani, Comp. Meth.
Diff. Eq. 1, 135 (2013)

\bibitem{ben9} F. Natali, Appl. Math. Lett. 23, 591 (2010)

\bibitem{ls01} G.M. Webb and G.P. Zank, J. Math. Phys. A: Math. Theor. 40,
545 (2007)

\bibitem{ls02} M.T.\ Mustafa, A.Y. Al-Dweik and R.A. Mara'beh, SIGMA 9, 041
(2013)

\bibitem{ls03} P.G.L. Leach, J. Math. Phys. 26, 510 (1985)

\bibitem{ls04} A.\ Paliathanasis and M.\ Tsamparlis, Int. J. Geom. Meth.
Mod. Phys. 11, 1450037 (2014)

\bibitem{ls05} S.V. Meleshko and V.P. Shapeev, J. Nonl. Math. Phys. 18, 195
(2011)

\bibitem{ls06} G.M. Webb, J. Phys A: Math. Gen. 23, 3885 (1990)

\bibitem{ls07} M. Tsamparlis and A. Paliathanasis, J. Phys. A: Math. Theor.
44, 175202 (2011)

\bibitem{pan2} P. Painlev\'{e}, Bulletin of the Mathematical Society of
France 28 201 (1900)

\bibitem{pan3} P. Painlev\'{e}, Acta Mathematica 25, 1 (1902)

\bibitem{pan4} P. Painlev\'{e}, Comptes Rendus de la Acad\'{e}mie des
Sciences de Paris 143, 1111 (1906)

\bibitem{kowa} S. Kowalevski, Acta. Math. 12, 177 (1889)

\bibitem{sin1} T. Brugarino and M.\ Sciacca, Optics Commun. 262, 250 (2006)

\bibitem{sin2} K. Andriopoulos and P.G.L. Leach, J. Math. Anal. Appl. 328,
860 (2007)

\bibitem{sin3} K. Andriopoulos and P.G.L. Leach, Appl. Anal. Disc. Math. 5,
230 (2011)

\bibitem{sin4} A. Paliathanasis, J.D. Barrow and P.G.L. Leach, Phys. Rev. D
94, 023525 (2016)

\bibitem{book1} G.W. Bluman and S. Kumei, Symmetries and Differential
Equations, Springer-Verlag, New York, (1989)

\bibitem{book2} P.J. Olver, Applications of Lie Groups to Differential
Equations, Springer-Verlag, New York, (1993)

\bibitem{book3} N.H. Ibragimov, CRC Handbook of Lie Group Analysis of
Differential Equations, Volume I: Symmetries, Exact Solutions, and
Conservation Laws, CRS Press LLC, Florida (2000)

\bibitem{bk04} M. Oberlack, M. Waclawczyk, A. Rosteck and V. Avsarkisov,
Mechanical Engineering Reviews, 2, 15-00157 (2015)

\bibitem{Abl1} M.J. Ablowitz, A. Ramani and H. Segur, \ Lettere al Nuovo
Cimento \textbf{23,} 333 (1978)

\bibitem{Abl2} M.J. Ablowitz, A. Ramani and H. Segur, J. Math. Phys. \textbf{%
21,} 715 (1980)

\bibitem{Abl3} M.J. Ablowitz, A. Ramani and H. Segur, J. Math. Phys. \textbf{%
21,} 1006 (1980)

\bibitem{buntis} A. Ramani, B. Grammaticos and T. Bountis, Physics Reports,
\textbf{180,} 159 (1989)

\bibitem{mor1} Morozov VV (1958), Izvestia Vysshikh Uchebn Zavendeni\u{\i}
Matematika, 5 161-171

\bibitem{mor2} Mubarakzyanov GM (1963), \textit{Izvestia Vysshikh Uchebn
Zavendeni\u{\i} Matematika,} \textbf{32} 114-123

\bibitem{mor3} Mubarakzyanov GM (1963), \textit{Izvestia Vysshikh Uchebn
Zavendeni\u{\i} Matematika,} \textbf{34} 99-106

\bibitem{mor4} Mubarakzyanov GM (1963), \textit{Izvestia Vysshikh Uchebn
Zavendeni\u{\i} Matematika,} \textbf{35} 104-116

\bibitem{olverb} P.J. Olver, Applications of Lie Groups to Differential
Equations, second edition, Springer-Verlag, New York (1993)

\bibitem{sinandronikos} A. Paliathanasis, A. Halder and P.G.L. Leach, A New
Viewpoint On Singularity Analysis, submitted (2019)

\bibitem{gbin} N.G. Berloff and L.H. Howard, Stud. Appl. Math. 99, 1 (1997)

\bibitem{ptpde1} R. Conte, Phys. Lett. A, 134, 100 (1988)

\bibitem{ptpde2} R. Conte, Phys. Lett. A, 140, 383 (1989)

\bibitem{ptpde3} M. Vlieg-Hulstman, Mathl. Comput. Modelling 18, 151 (1993)
\end{thebibliography}
\end{document}